\documentclass[square]{ws-procs975x65}
\usepackage{graphicx}

\newcommand{\vev}[1]{\langle #1\rangle}

\begin{document}

%\begin{flushright}
%VT-IPNAS-10-XX
%\end{flushright}

\title{Ratchet Model of Baryogenesis}

\author{Tatsu Takeuchi\footnote{Presenting author.}}
\address{%Institute for Particle, Nuclear, and Astronomical Sciences,\\
Physics Department, Virginia Tech, Blacksburg, VA 24061, USA}
\author{Azusa Minamizaki and Akio Sugamoto}
\address{Physics Department, Ochanomizu University, 2-1-1 \=Otsuka, Bunkyo-ku, Tokyo 112-8610, Japan}

\begin{abstract}
We propose a toy model of baryogenesis which applies 
the `ratchet mechanism,' used frequently in the theory of biological molecular motors, 
to a model proposed by Dimopoulos and Susskind.
\end{abstract}


\bodymatter

\section{Introduction}

The ratio of baryon-number to photon-number densities in our universe has been established
via Big-Bang Nucleosynthesis (BBN) \cite{Weinberg:1977ji,Steigman:2007xt,Fields:2010zz} and 
%the observation of the cosmic microwave background by 
WMAP \cite{Komatsu:2010fb,Fixsen:2009ug} to be
\begin{equation}
\eta \;=\; \dfrac{n_B}{n_\gamma} \;\approx\; 6\times 10^{-10}\;.
\end{equation}
The more precise numbers are
\begin{equation}
\begin{array}{ll}
\eta_{10}(\mathrm{BBN:D/H})
&\; = \;5.8\phantom{0}\pm 0.3\phantom{0}\;, \\
\eta_{10}(\mbox{WMAP: 7yr})
&\; = \;6.18\pm 0.15\;, 
\end{array}
\end{equation}
where $\eta_{10}=10^{10}\,\eta$, and the BBN value is determined from 
the deutron abundance reported in Ref.~\cite{O'Meara:2006mj,Pettini:2008mq}.
As we can see, the agreement is very good.

The objective of baryogenesis is to explain how the above number can come about
from a universe initially with zero net baryon number.
Since the pioneering work of Sakharov \cite{Sakharov},
very many proposals have been made as to what this baryogenesis mechanism could be.\footnote{For recent reviews, see Refs.~\cite{Riotto:1999yt,Dine:2003ax,Cline:2006ts,Buchmuller:2007fd,Shaposhnikov:2009zzb,Weinberg:2008zzc}.}
%Of these, one class of models
%generate baryon number via the coherent semi-classical
%time-evolution of a complex scalar field.
Among the early ones was a model by Dimopoulos and Susskind \cite{Dimopoulos:1978kv}
in which baryon number is generated via the coherent semi-classical time-evolution of
a complex scalar field.  
Similar mechanisms have been employed by Affleck and Dine \cite{Affleck:1984fy},
Cohen and Kaplan \cite{Cohen:1987vi}, and Dolgov and Freese \cite{Dolgov:1994zq}, 
of which the Affleck-Dine mechanism has been popular and intensely studied 
due to its natural implementability in SUSY models.
Two of us have also considered the application of the Dimopoulos-Susskind model
to the cosmological constant problem \cite{Minamizaki:2007kj}.

In this talk, I will discuss the Dimopoulos-Susskind model, 
how it satisfies Sakharov's three conditions for baryogenesis,
in particular, how it uses the expansion of the universe to satisfy the
third, and then propose the `ratchet mechanism' \cite{Reimann:1996,Julicher:1997zz,Reimann:2002} as an alternative for driving the model away from thermal equilibrium.

%%%%%%%%%%%%%%%%%%%%%%%%%%%%%%%%%%%%%%%%%%%%%%%%%%%%
\section{The Dimopoulos-Susskind Model}

Consider the action of a complex scalar field given by
\begin{equation}
S \;=\;
\int d^4x \,\sqrt{-g}\,
\Bigl[\;
g^{\mu\nu} \partial_\mu\phi^\dagger \partial_\nu\phi
- V(\phi,\phi^\dagger)
\,\Bigr] \;.
\end{equation}
If the potential $V(\phi,\phi^\dagger)$ is invariant under the global change of phase
\begin{equation}
\phi\;\rightarrow\;e^{i\xi}\phi\;,\qquad
\phi^\dagger\;\rightarrow\;e^{-i\xi}\phi^\dagger\;,
\end{equation}
then the corresponding conserved current is
\begin{equation}
B_\mu \;=\; \sqrt{-g}\Bigl(i\,\phi \stackrel{\leftrightarrow}{\partial_\mu}\!\phi^\dagger\Bigr)\;.
\end{equation}
If we identify $B_0$ with the baryon number density, then adding to the action
a potential which is not invariant under the above phase change, such as
\begin{equation}
V_0(\phi,\phi^\dagger) \;=\;
\lambda\, 
%\bigl(\phi^* \phi\bigr)^n 
\bigl(\,\phi+\phi^\dagger\bigr)
\bigl(\,\alpha\,\phi^3 + \alpha^*\phi^{\dagger 3}\bigr)\;,\qquad
|\alpha|\;=\;1\;,
\end{equation}
would lead to baryon number violation.
Furthermore, unless $\alpha = \pm 1$, 
this potential also violates $C$ and $CP$ since 
$\phi$ transforms as
\begin{eqnarray}
\phi(t,\vec{x})\;\; & \stackrel{C}{\longrightarrow}  &\;\; \phi^\dagger(t,\vec{x})\;,\cr
%\phi(t,\vec{x})\;\; & \stackrel{P}{\longrightarrow}  &\;\; \pm\phi\,(t,-\vec{x})\;,\cr
\phi(t,\vec{x})\;\; & \stackrel{CP}{\longrightarrow} &\;\; \pm\phi^\dagger(t,-\vec{x})\;,
\end{eqnarray}
where the sign under $CP$ depends on the parity of $\phi$. ($P$ is not violated.)

In Ref.~\cite{Dimopoulos:1978kv}, Dimopoulos and Susskind 
subject $\phi$ to the potential
\begin{equation}
V_n(\phi,\phi^\dagger) \;=\;
\lambda\, 
\bigl(\phi\,\phi^\dagger\bigr)^n 
\bigl(\,\phi+\phi^\dagger\bigr)
\bigl(\,\alpha\,\phi^3 + \alpha^*\phi^{\dagger 3}\bigr)
\;.
\end{equation}
The purpose of the factor $(\phi\,\phi^\dagger)^n$ is simply to give the coupling constant $\lambda$
a negative mass dimension.
Setting $\phi = \phi_r\,e^{i\theta}/\sqrt{2}$, the baryon number density becomes
\begin{equation}
n_B \;=\; B_0 \;=\; \sqrt{-g}\;\phi_r^2\,\dot{\theta}\;,
\end{equation}
which shows that to generate a non-zero baryon number $n_B$, one must generate a non-zero $\dot{\theta}$.
The potential in the polar representation of $\phi$ is
\begin{equation}
V_n(\phi_r,\theta)
\;=\;\lambda \left(\dfrac{\phi_r^{2}}{2}\right)^n \phi_r^4\,
\cos\theta\,\cos(3\theta+\beta)\;,
\label{Vtheta}
\end{equation}
where we have set $\alpha=e^{i\beta}$.
The $\theta$-dependence of this potential for fixed $\phi_r$ is shown in Fig.~1
for the case $\beta=\pi/2$.
\begin{figure}
\begin{center}
\includegraphics[width=8cm]{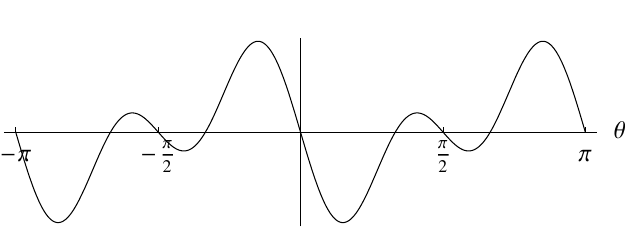}
\end{center}
\caption{$\theta$-dependence of the $B$, $C$, and $CP$ violating potential, Eq.~(\ref{Vtheta}), for the case
$\beta=\pi/2$.}
\end{figure}
Note that under $B$, $C$, and $CP$,
the phase $\theta$ transforms as 
\begin{eqnarray}
\theta(t,\vec{x}) \;\; & \stackrel{B}{\longrightarrow} & \;\; \theta(t,\vec{x})+\xi \;, \cr
\theta(t,\vec{x}) \;\; & \stackrel{C}{\longrightarrow} & \;\; -\theta(t,\vec{x}) \;, \cr 
\theta(t,\vec{x}) \;\; & \stackrel{CP}{\longrightarrow} & \;\; -\theta(t,-\vec{x}) \;.
\end{eqnarray}
If the parity of $\phi$ is negative, then $\theta$ will also be shifted by $\pi$
under $CP$.
So in terms of $\theta$, the violation of $B$ is due to the loss of translational invariance,
and the violation of $C$ and $CP$ are due to the loss of left-right reflection invariance
which happens when $\beta\neq 0,\pi$.
The question is, can the asymmetric force provided by this potential
make $\theta$ flow in one preferred direction thereby generate a non-zero 
$\dot{\theta}$?
For that, one must move away from thermal equilibrium.

In the original Dimopoulos-Susskind paper \cite{Dimopoulos:1978kv}, this
shift away from thermal equilibrium is accomplished by the expansion of the universe.
Consider a flat expanding universe with the Friedman-Robertson-Walker metric:
\begin{equation}
ds^2 \;=\; dt^2 - \left(\dfrac{a(t)}{a_0}\right)^2 d\vec{x}^2\;.
\end{equation}
%
%where $a_0=a(t_0)$ is the current scale of the universe.  
%Assuming that $\phi_r$ is constant, and $\theta$ is independent of $\vec{x}$, 
%The action of $\phi$ is
%
%\begin{equation}
%S \;=\;
%\int d^4x \,\sqrt{-g}\,
%\Bigl[\;
%g^{\mu\nu} \partial_\mu\phi^\dagger \partial_\nu\phi
%- V_n(\phi)
%\,\Bigr] \;,
%\end{equation}
%
%where
%
%\begin{equation}
%\sqrt{-g} \;=\; \left(\dfrac{a(t)}{a_0}\right)^3\;.
%\end{equation}
%
During a radiation dominated epoch, the scale factor evolves as
\begin{equation}
\dfrac{a(t)}{a_0}\;\sim\;\sqrt{2t}\;.
\end{equation}
Introducing the conformal variable $\tau=\sqrt{2t}$, the line-element
becomes
\begin{equation}
ds^2 \;=\; \tau^2\left(d\tau^2-d\vec{x}^2\right)\;,
\end{equation}
while the action simplifies to
\begin{equation}
S \;=\; \int d^3\vec{x}\,d\tau
\left[ \partial_\mu\hat{\phi}^\dagger\partial^\mu\hat{\phi}
-\dfrac{1}{\tau^{2n}}V_n(\hat{\phi},\hat{\phi}^\dagger)
+ \cdots
\right]\;.
\end{equation}
Here, the scalar field has been rescaled to $\hat{\phi}\;\equiv\;\tau\phi$,
and the ellipses represent total divergences and terms that depend only on $|\hat{\phi}|$.
%The baryon number density in the new variables is
%
%\begin{equation}
%n_B \;=\; B_0 \;=\; \hat{\phi}_r^2\,\dfrac{d\theta}{d\tau}\;.
%\end{equation}

At this point, a simplifying assumption is made that the dynamics of $|\hat{\phi}|$ is such that
it is essentially constant and does not evolve with $\tau$,
leaving only the phase of $\hat{\phi}$ as the dynamic variable.\footnote{
This assumption that $|\hat{\phi}|$ is constant would require
the magnitude of the unscaled field $|\phi|$ to evolve as $1/\tau = 1/\sqrt{2t}$.
%It is unclear whether such an evolution can be so conveniently realized.
}
Setting $\hat{\phi}= e^{i\theta}/\sqrt{2}$, the action
within a domain of spatially constant $\theta$ becomes
\begin{equation}
S \;=\; \int d^3\vec{x}\,d\tau
\left[ \dfrac{1}{2}\left(\dfrac{d\theta}{d\tau}\right)^2 
- \dfrac{1}{\tau^{2n}}V_n(1,\theta)
\right]\;.
\end{equation}
The equation of motion for $\theta$ within that domain is then
\begin{equation}
\dfrac{d^2\theta}{d\tau^2} + \dfrac{1}{\tau^{2n}}\dfrac{\partial V_n}{\partial\theta} \;=\; 0\;.
\end{equation}
To this, a friction term, which is assumed to come from the self-interaction of $\hat{\phi}$, 
is added by hand as
\begin{equation}
\dfrac{d^2\theta}{d\tau^2} 
+ \dfrac{1}{\tau^{2n}}\,\dfrac{\partial V_n}{\partial\theta} 
+ \dfrac{\lambda^2}{\tau^{4n}}\,\dfrac{d\theta}{d\tau} \;=\; 0\;,
\end{equation}
where the coefficient of $d\theta/d\tau$ has been fixed simply by dimensional analysis. 
If $n>0$, both force and friction terms vanish in the limit $\tau\rightarrow\infty$, 
and it is possible to show that a non-zero $n_B\sim d\theta/d\tau$ survives asymptotically, 
its final value depending on the initial value of $\theta$.
This initial value is expected to vary randomly from domain to domain, resulting in different
asymptotic baryon numbers in each, 
and when summed results in an overall net baryon number.
On the other hand, if $n=0$, which would make the self-interactions of $\phi$ 
renormalizable, the friction term will eventually bring all motion to a full stop.

%\begin{eqnarray}
%S & = &
%\int d^4x \,\sqrt{-g}\,
%\Bigl[\;
%g^{\mu\nu} \partial_\mu\phi^\dagger \partial_\nu\phi
%- V(\phi)
%\,\Bigr] \cr
%& \rightarrow & 
%\int d^4x \left(\dfrac{a(t)}{a_0}\right)^3
%\left[ \dfrac{1}{2}\,\phi_r^2\,\dot{\theta}^2 - V(\phi_r,\theta)
%\right]
%\;,
%\end{eqnarray}
%
%The presence of $\sqrt{-g}$ modifies the baryon number density to
%
%\begin{equation}
%n_B 
%\;=\; \sqrt{-g}\,i\,\phi \stackrel{\leftrightarrow}{\partial_0}\!\phi^\dagger
%\;=\; \left(\dfrac{a(t)}{a_0}\right)^3\phi_r^2\,\dot{\theta}\;.
%\end{equation}
%
%and the equation of motion of $\theta$ is
%
%\begin{equation}
%\ddot{\theta} + 3H\,\dot{\theta}+\dfrac{1}{\phi_r^2}\dfrac{\partial V}{\partial\theta} \;=\; 0\;,
%\end{equation}
%
%where
%
%\begin{equation}
%H 
%\;=\; \dfrac{\dot{a}(t)}{a(t)}
%\;=\; 
%\left\{ \begin{array}{ll}
%\dfrac{1}{2t}\qquad\qquad &\mbox{radiation dominated} \\
%\phantom{-} & \\
%\dfrac{2}{3t}\qquad  &\mbox{matter dominated} 
%\end{array}
%\right.
%\;.
%\end{equation}
%

%%%%%%%%%%%%%%%%%%%%%%%%%%%%
\section{The Ratchet Mechanism}

A striking feature of the Dimopoulos-Susskind model is its similarity with the
problem of biased random walk one encounters in the modeling of biological motors
\cite{Reimann:1996,Julicher:1997zz,Reimann:2002}.
An example of a biological motor is the myosin molecule which walks along
actin filaments.  This molecule is modeled as moving along a periodic
sawtooth-shaped potential, similar to that shown in Fig.~1.
Thermal equilibrium inside a living organism is broken by the presence of ATP 
(adenosine triphosphate) whose
hydrolysis into ADP (adenosine diphosphate) and P (phosphate)
provides the energy required to fuel the motion:
\begin{equation}
\mathrm{ATP} \;\rightarrow\; \mathrm{ADP} + \mathrm{P} + \mbox{energy}\;.
\end{equation}
This is often modeled as a randomly fluctuating temperature of the thermal bath:
the molecule is excited out of a potential well during periods of high-temperature, 
allowing it to diffuse into the neighboring ones, and then drops back into a well 
during periods of low-temperature.  Due to the asymmetry of the potential, this sequence
can lead to biased motion depending on the depth and width of the repeating potential wells, 
and the height and frequency of the temperature fluctuations.

Analogy with such `temperature ratchet' models 
suggests a possible way to drive the evolution of $\theta$ in the
Dimopoulos-Susskind model without relying on the non-renormalizability of the self-interaction of $\phi$,
or the expansion of the universe directly.
Let us assume the existence of ATP- and ADP-like particles $A$ and
$B$ which interact with $\phi$ via the reaction
\begin{equation}
A + \phi \;\leftrightarrow\; B + \phi + Q\;,
\end{equation}
where $Q$ is the energy released in the reaction.  $A$ and $B$ are assumed to 
be stable (or highly meta-stable) states that have fallen out of thermal equilibrium at 
an earlier time in the evolution of the universe.  Though they interact with $\phi$,
giving or taking energy away from it, their masses are such that the decay
\begin{equation}
A\rightarrow B + \phi + \bar{\phi}
\end{equation}
is kinematically forbidden. 

In order to isolate the effect of the presence of a bath of these particles,
we neglect the expansion of the universe and 
subject $\phi=\phi_r\,e^{i\theta}/\sqrt{2}$ 
to the $n=0$ renormalizable Dimopoulos-Susskind potential $V_0(\phi_r,\theta)$.
We again adopt the simplifying assumption that the evolution of $\phi_r$ is suppressed.
%
%where the ellipses represent unspecified terms that depend only on $|\phi|$ that stabilize
%$V_n(\phi,\phi^\dagger)$.
%which renders the potential for $|\phi|$ 
%into a mexican-hat-like potential which gives the radial excitation 
%of $\phi$ a large enough mass around a non-zero vacuum expectation value
%to effectively decouple it and leave only the phase
%of $\phi$ dynamical.\footnote{Alternatively, $|\phi|$ could evolve very slowly compared
%to the evolution of $\theta$ so that it is effectively constant.}
%
Though the interactions between $\phi$ and the $A$ and $B$ particles occur randomly,
we model their effect by a periodically fluctuating kinetic energy of $\theta$ \cite{Reimann:1996} :
\begin{equation}
K(t) \;=\; K_0 \Bigl[ 1+ A\sin(\omega t)\Bigr]^2\;.
\end{equation}
This function oscillates between $K_\mathrm{min}=K_0(1-A)^2$ and $K_\mathrm{max}=K_0(1+A)^2=K_\mathrm{min}+Q$.
Therefore,
\begin{equation}
Q \;=\; 4K_0 A\;.
\end{equation}
Then, the equation of motion of $\theta$ in our model will be given by the Langevin equation
\begin{equation}
\phi_r^2\,\ddot{\theta} \;=\;
-\dfrac{\partial V_0}{\partial\theta}
-\eta\,\dot{\theta} + \sqrt{4\eta K(t)}\,\xi(t)\;,
\end{equation}
where $\eta$ is the coefficient of friction, and $\xi(t)$ is Gaussian white noise:
\begin{equation}
\vev{\xi(t)}\;=\;0\;,\qquad
\vev{\xi(t)\xi(s)}\;=\;\delta(t-s)\;.
\end{equation}
The above Langevin equation is equivalent to the following Fokker-Planck equation governing the evolution
of the probability density $p(\theta,t)$ and the probability current $j(\theta,t)$:
\begin{eqnarray}
0 & = & \dfrac{\partial p(\theta,t)}{\partial t} + \dfrac{\partial j(\theta,t)}{\partial\theta}\;,\cr
j(\theta,t) & = & -\dfrac{1}{\eta}\left[
\dfrac{\partial V_0}{\partial\theta}\,p(\theta,t)
+ 2K(t)\dfrac{\partial p(\theta,t)}{\partial \theta}
\right]\;.
\end{eqnarray}
The quantity of interest for baryon number generation is the period-averaged probability current
\begin{equation}
J \;=\; \dfrac{1}{\mathcal{T}}\int_0^\mathcal{T} j(\theta,t)\,dt
%\;\sim\; \vev{\dot{\theta}}
\;,
\end{equation}
which is asymptotically independent of $\theta$ and approaches a constant,
a non-zero value signifying a non-zero baryon number.
For the sake of simplicity, we set $\beta=\pi/2$, and
$\phi_r$, $\lambda$, and $\eta$ all equal to one.
We then solved these equations numerically 
for various values of $K_\mathrm{min}$, $\omega$, and $Q$,
and have found that non-zero $J$ can be generated for a very wide range of parameter choices.
As an example, we show the $\omega$-dependence of $J$ for the case $K_\mathrm{min}=0.5$ and $Q=1$
in Fig.~2.
Further details of our analysis can be found in Ref.~\cite{Minamizaki:2010}.
\begin{figure}[t]
\begin{center}
\includegraphics[width=7cm]{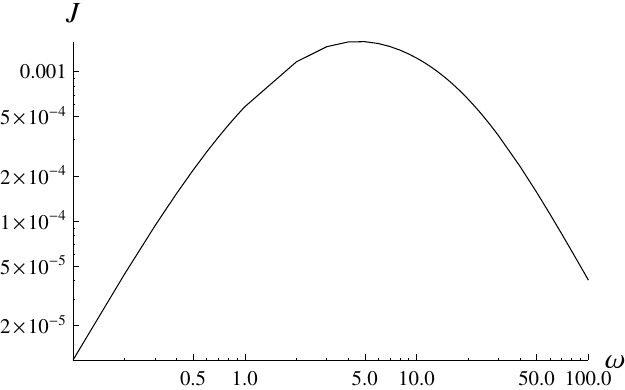}
\end{center}
\caption{$\omega$-dependence of $J$ for the case
$\beta=\pi/2$, $K_\mathrm{min}=0.5$, $Q=\eta=\phi_r=\lambda=1$.}
\end{figure}
%

%%%%%%%%%%%%%%%%%%%%%%%%%%%%%%%%%%%%%%%%%%
\section{What is the ATP-like particle?}

Whether the ratchet mechanism we are proposing here can be embedded into a realistic
scenario remains to be seen.
Of particular difficulty may be maintaining a sufficiently large population of the ATP-like particles to
drive the ratchet.
But what can these ATP-like particles be?
Several possibilities come to mind:
First, it could be the inflaton at reheating, transferring energy to the $\phi$ field
via parametric resonance. 
Second, they could be heavy Kaluza-Klein (KK) modes in some extra-dimension model.
And third, perhaps they could be technibaryons transferring energy to technimeson $\phi$'s.
Finally, regardless of what their actual identities are, if the ATP-like particles are highly stable and still around, they may constitute dark matter, thereby connecting baryogenesis with the dark matter problem.
These, and other possibilities will be discussed elsewhere \cite{MST:2010}.

%\newpage

\section*{Acknowledgments}

We would like to thank Philip Argyres and Daniel Chung for
helpful suggestions. 
T.T. is supported by the U.S. Department of Energy, 
grant DE--FG05--92ER40709, Task A.

\end{document}